\documentclass[%
 aip,
 amsmath,amssymb,
 reprint,%
 floatfix,
]{revtex4-1}

\usepackage{graphicx}% Include figure files
\usepackage{dcolumn}% Align table columns on decimal point
\usepackage{bm}% bold math
\usepackage[utf8]{inputenc}
\usepackage[T1]{fontenc}
\usepackage{mathptmx}
\usepackage{etoolbox}
\usepackage{color}

\makeatletter
\def\@email#1#2{%
 \endgroup
 \patchcmd{\titleblock@produce}
  {\frontmatter@RRAPformat}
  {\frontmatter@RRAPformat{\produce@RRAP{*#1\href{mailto:#2}{#2}}}\frontmatter@RRAPformat}
  {}{}
}%
\makeatother
\begin{document}

\preprint{AIP/123-QED}

\title[Insensitivity of a turbulent laser-plasma dynamo to initial conditions]{Insensitivity of a turbulent laser-plasma dynamo to initial conditions}
% Force line breaks with \\
\author{A.F.A. Bott}
 \email{abott@princeton.edu}
\affiliation{ Department of Physics, University of Oxford, Oxford, UK}
\affiliation{Department of Astrophysical Sciences, University of Princeton, Princeton, USA}

\author{L. Chen}
\affiliation{ Department of Physics, University of Oxford, Oxford, UK}

\author{P. Tzeferacos}%
\affiliation{ Department of Physics, University of Oxford, Oxford, UK}
\affiliation{Department of Physics and Astronomy, University of Rochester, Rochester, USA}
\affiliation{Laboratory for Laser Energetics, University of Rochester, Rochester, USA}

\author{C.A.J. Palmer}
\affiliation{School of Mathematics and Physics, Queens University Belfast, Belfast, UK}

\author{A.R. Bell}
\affiliation{Department of Physics, University of Oxford, Oxford, UK}

\author{R. Bingham}
\affiliation{Rutherford Appleton Laboratory, Chilton, Didcot, UK}
\affiliation{Department of Physics, University of Strathclyde, Glasgow, UK}

\author{A. Birkel}
\affiliation{Massachusetts Institute of Technology, Cambridge, USA}

\author{D.H. Froula}
\affiliation{Department of Physics and Astronomy, University of Rochester, Rochester, USA}
\affiliation{Laboratory for Laser Energetics, University of Rochester, Rochester, USA}

\author{J. Katz}
\affiliation{Laboratory for Laser Energetics, University of Rochester, Rochester, USA}

\author{M.W. Kunz}
\affiliation{Department of Astrophysical Sciences, University of Princeton, Princeton, USA}

\author{C.-K. Li}
\affiliation{Massachusetts Institute of Technology, Cambridge, USA}

\author{H-S. Park}
\affiliation{Lawrence Livermore National Laboratory, Livermore, USA}

\author{R. Petrasso}
\affiliation{Massachusetts Institute of Technology, Cambridge, USA}

\author{J.S. Ross}
\affiliation{Lawrence Livermore National Laboratory, Livermore, USA}

\author{B. Reville}
\affiliation{Max-Planck-Institut für Kernphysik, Heidelberg, Germany }

\author{D. Ryu}
\affiliation{Department of Physics, School of Natural Sciences, UNIST, Ulsan, Korea}

\author{F.H. S\'eguin}
\affiliation{Massachusetts Institute of Technology, Cambridge, USA}

\author{T.G. White}
\affiliation{Department of Physics, University of Nevada, Reno, Nevada 89557, USA}

\author{A.A. Schekochihin}
\affiliation{ Department of Physics, University of Oxford, Oxford, UK}

\author{D.Q. Lamb}
\affiliation{Department of Astronomy and Astrophysics, University of Chicago, Chicago, USA}

\author{G. Gregori}
\affiliation{ Department of Physics, University of Oxford, Oxford, UK}
\affiliation{Department of Astronomy and Astrophysics, University of Chicago, Chicago, USA}

\date{\today}% It is always \today, today,

\begin{abstract}
It has recently been demonstrated experimentally that a turbulent plasma created by the collision of two inhomogeneous, asymmetric, weakly magnetised laser-produced plasma jets can generate strong stochastic magnetic fields via the small-scale turbulent dynamo mechanism, provided the magnetic Reynolds number of the plasma is sufficiently large. In this paper, we compare such a plasma with one arising from two pre-magnetised plasma jets whose creation is identical save for the addition of a strong external magnetic field imposed by a pulsed magnetic field generator (`MIFEDS'). We investigate the differences between the two turbulent systems using a Thomson-scattering diagnostic, X-ray self-emission imaging and proton radiography. The Thomson-scattering spectra and X-ray images suggest that the presence of the external magnetic field has a limited effect on the plasma dynamics in the experiment. While the presence of the external magnetic field induces collimation of the flows in the colliding plasma jets and the initial strengths of the magnetic fields arising from the interaction between the colliding jets are significantly larger as a result of the external field, the energy and morphology of the stochastic magnetic fields post-amplification are indistinguishable. We conclude that, for turbulent laser-plasmas with super-critical magnetic Reynolds numbers, the dynamo-amplified magnetic fields are determined by the turbulent dynamics rather than the seed fields and modest changes in the initial flow dynamics of the plasma, a finding consistent with theoretical expectations and simulations of turbulent dynamos.
\end{abstract}

\maketitle

\section{\label{sec:intro}Introduction}

Explaining the dynamically significant magnetic fields that are routinely observed in various astrophysical environments composed of plasma (such as the intracluster medium~\cite{Vacca2018}) is a problem that has occupied astrophysicists for over half a century~\cite{Biermann1951}. One of the most plausible mechanisms that can account for the strength of such magnetic fields is the small-scale turbulent dynamo, whereby turbulent bulk motions of a magnetised fluid or plasma cause the efficient amplification of magnetic fields until they possess energies that are comparable to the kinetic energy of the driving turbulent motions~\cite{Batchelor1950,Ryu2008}. A significant number of analytical calculations~\cite{Kazantsev1968,Vainstein1972,Zeldovich1984,Kulsrud1992} and simulations~\cite{Meneguzzi1981,Kida1991,Miller1996,Cho2001,Schekochihin2004,Haugen2004,Schekochihin2007,Cho2009,Beresnyak2012,Porter2012,Seta2020,Seta2020b} support the efficacy of this mechanism provided the magnetic Reynolds number $\mathrm{Rm}$ of the plasma (defined as $\mathrm{Rm} \equiv u_{\rm rms} L/\eta$, where $u_{\rm rms}$ is the root-mean-square (rms) magnitude of the turbulent motions, $L$ is the characteristic scale of those motions, and $\eta$ the plasma’s resistivity) exceeds some critical value $\mathrm{Rm}_{\rm c}\approx50$-$400$~\cite{Rincon2019}. 

Of particular importance is the expectation that the characteristic strength of the magnetic field post-amplification depends only on the turbulent kinetic energy, and is insensitive to both the strength of initial seed magnetic fields and subtle particularities of the turbulent flow dynamics pre-amplification, provided Rm is supercritical (viz., $\mathrm{Rm} > \mathrm{Rm}_{\rm c}$)~\cite{Schekochihin2007,Seta2020}. 
{\color{black}This expectation arises because the induction equation that is thought to describe the magnetic field's evolution is linear in the magnetic field, and so the saturation of dynamo-amplified  fields can only be enforced by the back-reaction of the Lorentz force on the turbulent flow dynamics. This saturation mechanism sets the precise strengths at which magnetic fields are maintained, a quantity of great significance in astrophysical contexts. It also} allows amplification of the initial magnetic energy over many orders of magnitude if it is much smaller than the turbulent kinetic energy.
In many astrophysical environments, this property is crucial for resolving the vast discrepancy between the characteristic magnitudes of the observed dynamical fields, and seed magnetic fields generated by processes that can produce magnetic fields in unmagnetized plasmas (such as the Biermann battery~\cite{Biermann1951,Kulsrud1997}).

In the last two decades, it has become possible to explore dynamo processes in controlled laboratory experiments. Historically, the first such experiments involved liquid-metal flows, which yielded many significant results: the first kinematic dynamo flow~\cite{Gailitis2000}, dynamo saturation~\cite{Gailitis2001}, and dynamo action in a partially stochastic flow~\cite{Monchaux2007}. However, liquid-metal experiments are limited to certain parameter regimes: incompressible flows whose magnetic Prandtl number $\mathrm{Pm}$ (defined as $\mathrm{Pm} \equiv \nu/\eta$, where $\nu$ is the fluid viscosity) is much smaller than unity. Since $\mathrm{Pm}$ is known to be an important parameter for the operation of turbulent dynamos~\cite{Schekochihin2007}, and is large in many astrophysical environments~\cite{Schekochihin2004}, alternative experimental approaches are needed. A series of recent laser-plasma experiments have started to satisfy this need, producing a series of notable results: first the demonstration of the amplification of seed magnetic fields generated by the Biermann battery~\cite{Gregori2012,Meinecke2014,Meinecke2015,Gregori2015}, and then the operation of a \emph{bona fide} small-scale turbulent dynamo in a plasma with $\mathrm{Rm} \approx 600$~\cite{Tzeferacos2017,Tzeferacos2018}. In the last year, another laser-plasma experiment provided time-resolved measurements of the action of a small-scale turbulent dynamo with $\mathrm{Rm} \approx 450$, and also accessed the $\mathrm{Pm}\sim 1$ regime for the first time in the laboratory~\cite{Bott2021} thanks to design improvements to the platform~\cite{Muller2017}. 
Most recently, amplification of magnetic fields (but not yet dynamo) was observed in a supersonic turbulent laser-plasma~\cite{Bott2021b}, an experiment that was based on the first successful realisation of boundary-free supersonic turbulence in the laboratory~\cite{White2019}.  

One finding of previous turbulent-dynamo experiments
that merited further study was the ratio of the magnetic to turbulent kinetic energy-density, which was observed to be finite, but still quite small ($\epsilon_{\rm B}/\epsilon_{\rm K,turb} \approx3$-$4\%$)~\cite{Tzeferacos2018,Bott2021}. 
This prompted the question of whether the characteristic post-amplification strength of magnetic fields in these turbulent laser-plasmas is determined by the plasma's turbulent kinetic energy alone, or was in fact not fully saturated and thus might be expected to be increased by a stronger initial magnetic field.
In this paper, we discuss results from a new experiment (in the $\mathrm{Pm} \sim 1$ regime) on the Omega Laser Facility~\cite{Boehly1997} that demonstrates this claim. This demonstration is made by introducing a seed magnetic field into a turbulent laser-plasma with $\mathrm{Rm} > \mathrm{Rm}_{\rm c}$ whose energy is over an order of magnitude larger than the energy of the seed field arising inherently in the plasma.
The stochastic magnetic fields that arise from the action of the small-scale turbulent dynamo on the seed field are then compared with a control case (viz., a plasma without an enhanced seed field). The key result is that the characteristic strengths and structure of the magnetic fields are indistinguishable with or without the enhanced seed field and its modest effect on the initial flow dynamics of the plasma.

\section{Experimental design} \label{sec:exp_design}

\begin{figure}
\includegraphics[width=\linewidth]{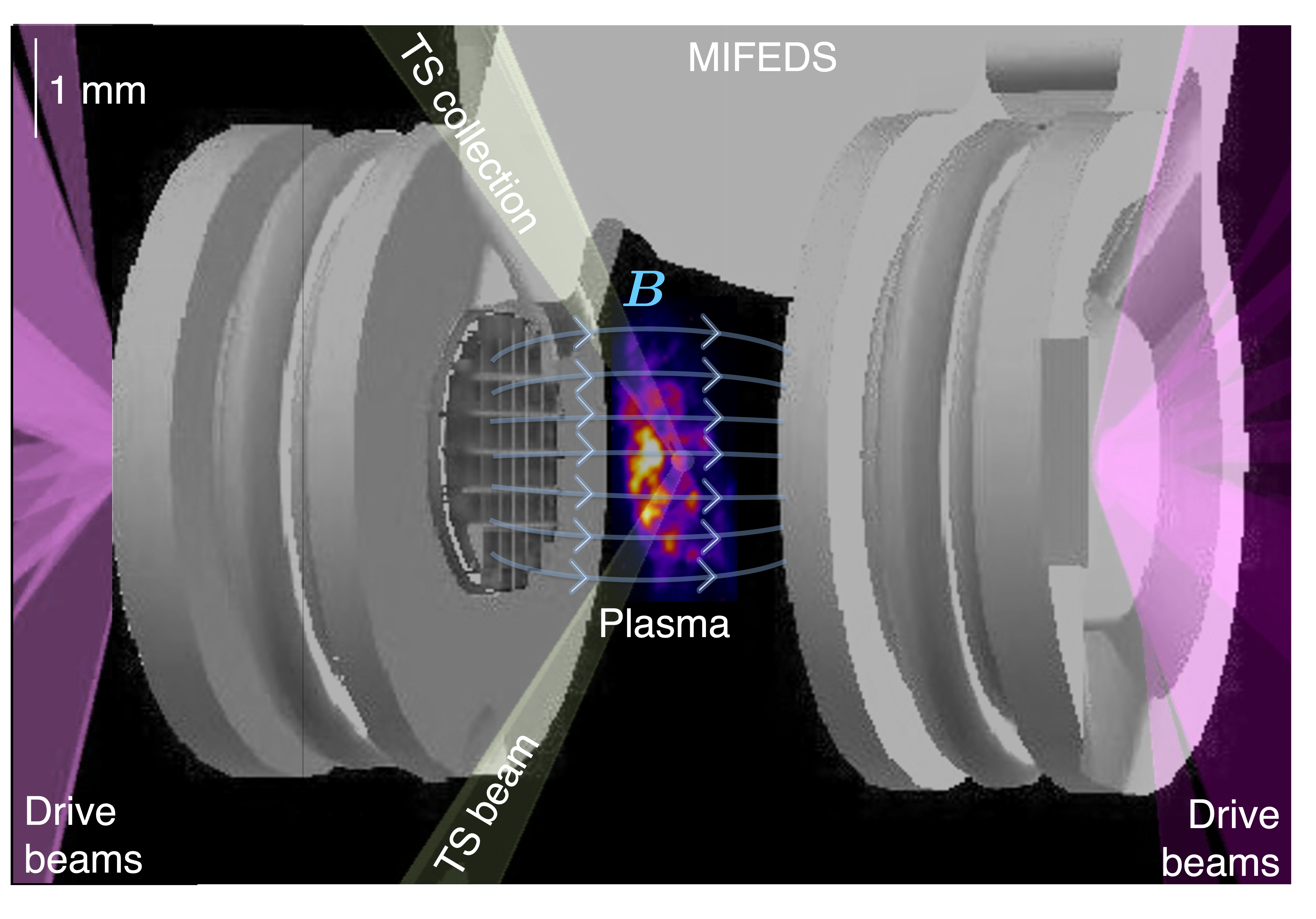}
\caption{\label{fig:expdesign} \emph{Experimental set-up.} Schematic of the experimental platform. Twenty beams of the OMEGA laser deliver a total of 10 kJ of 351-nm wavelength laser-light energy over 10 ns to an 800-µm diameter focal spot on two CH plastic foils (5 kJ/foil). The foils have the same design used on a previous experiment on the Omega Laser Facility~\cite{Bott2021}: the primary foils have thicknesses of 50 µm and are attached to annular washers with outer diameter of 3 mm, central-hole diameter of 400 µm, and a thickness of 230 µm. The grids used on the target are also the same; they have 300 µm square holes and 100 µm wires, and are asymmetric, with the midpoints of the holes in one grid always facing the midpoints of the wire intersections in the other. The MIFEDS is operated at 19 kV, with maximum voltage coinciding with drive-beam initiation; the morphology of the initial magnetic field is indicated in light blue. The location of the interaction-region plasma is indicated, as is the Thomson scattering probe beam's path.}
\end{figure}

The schematic of the experimental platform is shown in Figure \ref{fig:expdesign}, with detailed target specifications given in the caption. 
The target platform design is related to previous experiments on the Omega Laser Facility~\cite{Tzeferacos2018, Chen2020, Bott2021}: a turbulent plasma is created by colliding rear-side blow-off plasma jets that, prior to their collision, have passed through asymmetric grids. On collision, an `interaction region’ of plasma forms which has higher characteristic densities and ion/electron temperatures than either jet. In addition, the asymmetric heterogeneity of the jets leads to the formation of strong shear flows in the interaction-region plasma. These become Kelvin-Helmholtz unstable and turbulent motions quickly develop.

The experimental platform outlined here differs from previous experiments in one key regard: the whole target assembly is embedded inside a pulsed magnetic-field generator (the ‘magneto-inertial fusion electrical discharge system’, or MIFEDS~\cite{Gotchev2009,Fiksel2015}). When utilised, the MIFEDS generates a magnetic field with a magnitude of ${\sim}$150 kG at the target foil and ${\sim}$80 kG at the target’s centre that is oriented approximately parallel to the axis which passes through the geometric centres of the foils and grids, and also along which both plasma jets propagate (`the line of centres'); Figure 1 shows a schematic of the magnetic field lines generated by the MIFEDS between the two grids. The magnitude of this magnetic field is significantly larger than that of the magnetic fields (${\sim}$10 kG) generated by the Biermann battery and advected to the target's midpoint by the jets~\cite{Bott2021}. However, the effect of a magnetic field of this strength on the dynamics of the interaction-region plasma is modest (as we explicitly demonstrate in sections \ref{sec:Xray} and \ref{sec:TS}).
Thus, this platform allowed us to test whether introducing a much larger seed magnetic field into the interaction-region plasma changes the magnitude of magnetic fields amplified by turbulent motions.

We characterised the turbulent plasma in both the presence and absence of the MIFEDS (which we refer to as `MIFEDS experiments' and `no-MIFEDS experiments', respectively) using three laser-plasma diagnostics: self-emission X-ray imaging to diagnose the plasma's dynamical evolution and turbulence, a time-resolved Thomson-scattering diagnostic to assess the plasma’s physical state, and proton radiography using a D$^{3}$He backlighter capsule to characterise magnetic fields. {\color{black} The set-up of all of these diagnostics was similar to that used in previous turbulent-dynamo experiments on the Omega Laser Facility~\cite{Tzeferacos2018,Bott2021}, as was our methodology for analysing the data that was collected from them. However, for the sake of clarity and completeness, we provide a self-contained exposition here for each diagnostic that both reviews our approach and notes the aspects that are unique to this experiment.}

\section{Results}

\subsection{Characterising turbulence: X-ray self-emission imaging} \label{sec:Xray}

\begin{figure}
\includegraphics[width=\linewidth]{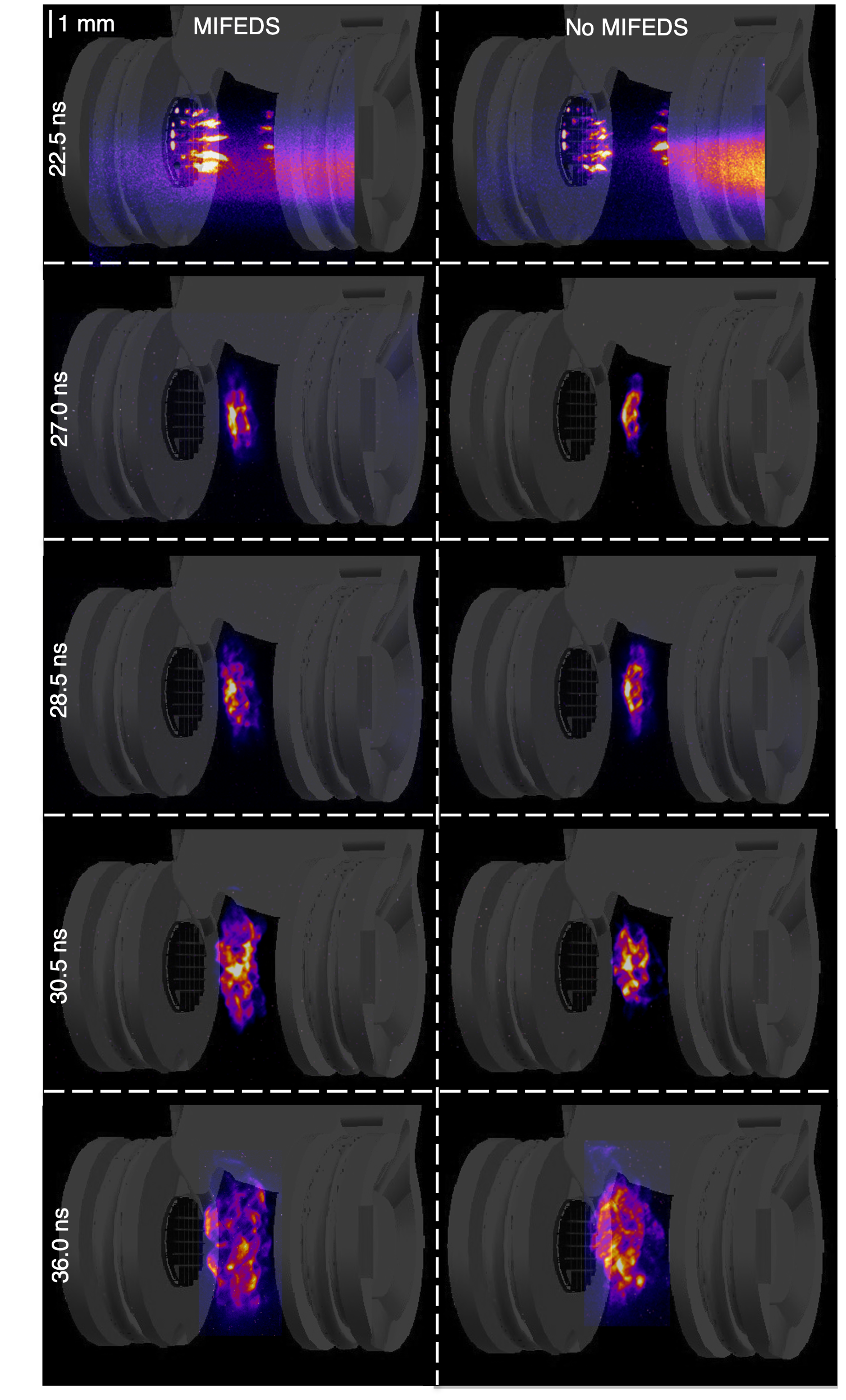}
\caption{\label{fig:Xrayimages} \emph{X-ray self-emission images.} XRFC images of soft X-rays emitted by the turbulent plasma in both the presence (left) and absence (right) of the MIFEDS. The top row (22.5 ns after the start of the drive-beam) employed a 100 V bias on the XFRC, while all other (post-collision) images used 350 V (the former having 32$\times$ sensitivity). The resolution of the images, which is set by the pinhole size and the MCP's response, is ${\sim}$50 µm. For reference, a projection of the target is shown on each image as a gray shade.}
\end{figure}

{\color{black} The X-ray imaging diagnostic measured soft X-rays (${\sim}0.2$-$0.5$ keV) emitted by free-free bremsstrahlung in the fully ionized CH plasma using a pinhole X-ray framing camera (XRFC) configured with a two-strip microchannel plate (MCP) and charged-coupled device (CCD) camera at a range of different times~\cite{Kilkenny1988,Bradley1995}.} The technical specifications of this XRFC were identical to that of the previous experiments~\cite{Bott2021}; the magnification of the imaging was 2$\times$, the pinhole diameter was 50 µm, a thin filter (0.5 µm polypropene with a 150 nm aluminium coating) was positioned in front of the MCP to block low-energy ($\lesssim 0.1$ keV) electromagnetic radiation, and each strip of the MCP was operated at two independent times, each with a 1 ns gate. The only difference with previous X-ray imaging diagnostic set-ups was the orientation of the camera; in this experiment, the XRFC was oriented at a ${\sim}30^{\circ}$ angle with respect to the plane of the interaction-region plasma (60$^{\circ}$ with respect to the line of centres), in order to observe fluctuations in the plasma's emission within that plane. Previously, X-ray imaging was carried out in a side-on configuration  (viz., at 90$^{\circ}$ with respect to the line of centres); however, on account of the interaction region's narrow extent with respect to the line of centres for a ${\sim}$5 ns interval subsequent to collision, detecting turbulent fluctuations during this time interval using this configuration proved to be challenging.  XRFC images from the experiment both before and after the jet collision are shown in Figure \ref{fig:Xrayimages}. 

\begin{figure}
\includegraphics[width=\linewidth]{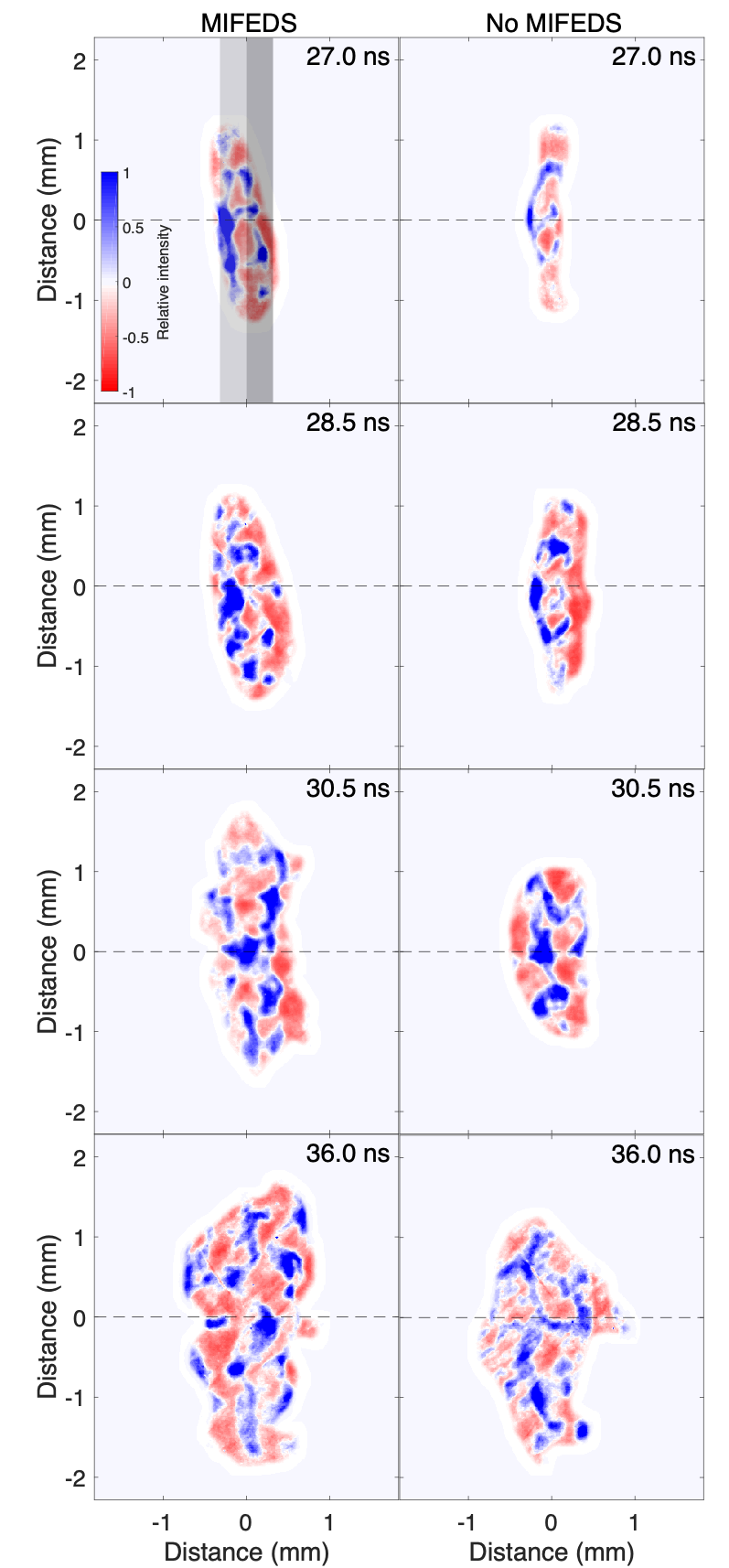}
\caption{\label{fig:Xrayimages_rel} \emph{Relative-intensity maps.} Maps of fluctuations in the detected X-ray intensity relative to a smooth mean intensity profile. The mean profiles are determined from the raw data using a two-dimensional $81 \times 81$ pixel mean filter (which, given the effective 9 µm pixel size of each images, corresponds to a characteristic scale of ${\sim}$0.8 mm), which is combined with a two-dimensional Gaussian window function (characteristic scale 150 µm). The window function is utilised in order to account for the sharp drop in the measured X-ray intensity associated with the accretion shocks that circumscribe the interaction-region plasma. The gray shaded regions denote the intervals over which the mean X-ray intensity profile is averaged when determining the transverse extent of the interaction-region plasma (see text).}
\end{figure}

Before the collision in both the MIFEDS and no-MIFEDS experiments (top row of Figure \ref{fig:Xrayimages}), we observe finger-like regions of emission from the plasma jets. Once the collision between these jets has occurred (second and subsequent rows of Figure \ref{fig:Xrayimages}) a region of strong, fluctuating emission (originating from the interaction-region plasma) develops. The fluctuations are related to density inhomogeneities in the plasma, whose origin can in turn be attributed to the effect of turbulent motions. Using a technique that was previously applied to similar X-ray imaging data~\cite{Bott2021}, we can extract `maps' of relative-intensity fluctuations for each of these post-collision images by first constructing a smooth mean X-ray intensity profile, and then dividing the total intensity by this profile; the resulting relative-intensity maps are shown in Figure \ref{fig:Xrayimages_rel}. 

Comparing the X-ray images obtained in the MIFEDS and no-MIFEDS experiments, we find the emission from the incident plasma jets to be slightly more extended and collimated in the former case (a physical explanation for this effect is provided with the help of Thomson scattering data in Section \ref{sec:TS}). 
However, once the interaction-region plasma has formed,
the emission is qualitatively similar irrespective of whether the MIFEDS is turned on or not -- this applies to
both the size of the region from which X-rays are emitted and the fluctuations in the X-ray intensity. 

\begin{figure}[htbp]
\includegraphics[width=\linewidth]{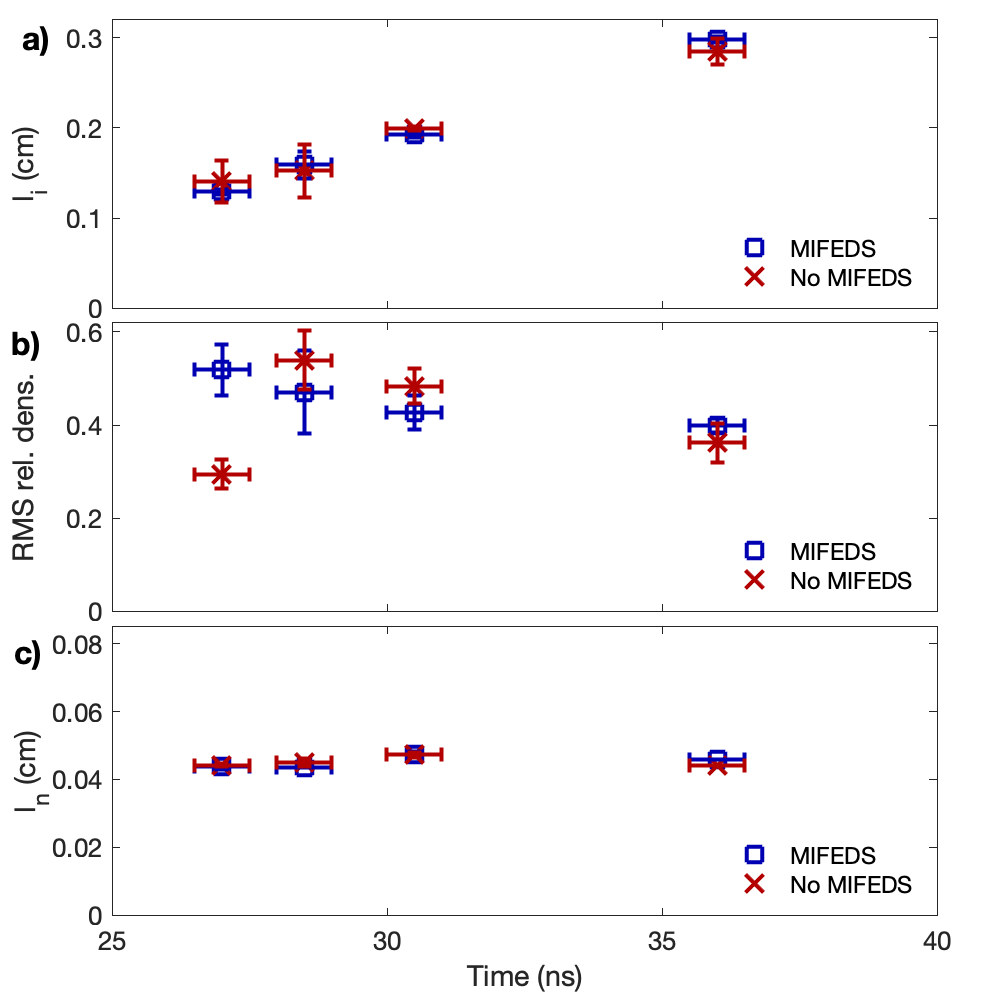}
\caption{\label{fig:Xrayimages_quant} \emph{Quantitative analysis of X-ray images.} \textbf{a)} Evolution of the interaction-region transverse width $l_i$ over time in the presence (blue squares) and absence (red crosses) of MIFEDS. The uncertainty of the measurement is estimated by assuming that the interaction-region plasma is approximately homogeneous, and then considering the left-hand and right-hand sides of the interaction region as independent samples (cf. Figure \ref{fig:Xrayimages_rel}, top left panel). \textbf{b)} Evolution over time of the inferred rms magnitude of density fluctuations in the plasma. The uncertainty of the measurement is estimated in a similar manner to that described for a), but using the upper and lower regions of the interaction-region plasma instead of the left- and right-hand sides. \textbf{c)} Evolution over time of the inferred integral length $l_n$ of density fluctuations in the plasma. The uncertainty is calculated in the same way as described for b).}
\end{figure}

To confirm these conclusions more robustly, we perform a quantitative analysis on the X-ray images. First, we measure 
the transverse extent $l_i$ of the interaction-region plasma in both MIFEDS and no-MIFEDS experiments by averaging the same mean X-ray intensity profiles that we mentioned previously in the direction parallel to the line of centres, and then by calculating the full-width-half-maximum (FWHM) of the resulting one-dimensional profile. The results, which are presented in Figure \ref{fig:Xrayimages_quant}a, show that $l_i$ in both types of experiment is indeed indistinguishable within the error of the measurement.
To show quantitatively that the statistical properties of the turbulence
are not significantly affected by the presence of the MIFEDS, we use the fact that the interaction-region plasma is optically thin to its bremsstrahlung-dominated X-ray emission~\cite{Bott2021} to relate the relative intensity fluctuations to path-integrated relative density fluctuations~\cite{Churazov2012}. Then, under the assumption of approximately isotropic and homogeneous density statistics (assumptions justified by previous analysis of similar experiments~\cite{Bott2021}), we can determine the rms of relative density fluctuations and their integral scale $l_n$. These quantities are shown in Figures \ref{fig:Xrayimages_quant}b and \ref{fig:Xrayimages_quant}c, respectively. The results are again similar for the MIFEDS and no-MIFEDS experiments, with one exception: the rms of relative density fluctuations not long after collision is larger in the presence of the MIFEDS magnetic field. We attribute this difference to a slightly earlier collision time in the MIFEDS experiments (see section \ref{sec:TS}) allowing for earlier onset of turbulent motions. We note that the characteristic value (${\sim}0.5$) of the rms of relative density fluctuations in both the MIFEDS and no-MIFEDS experiments is close to values derived in previous experiments, as is the value of the integral scale $l_n$ which is comparable to the grid periodicity $L \approx 400$ µm~\cite{Bott2021}. 

\begin{figure}[htbp]
\includegraphics[width=\linewidth]{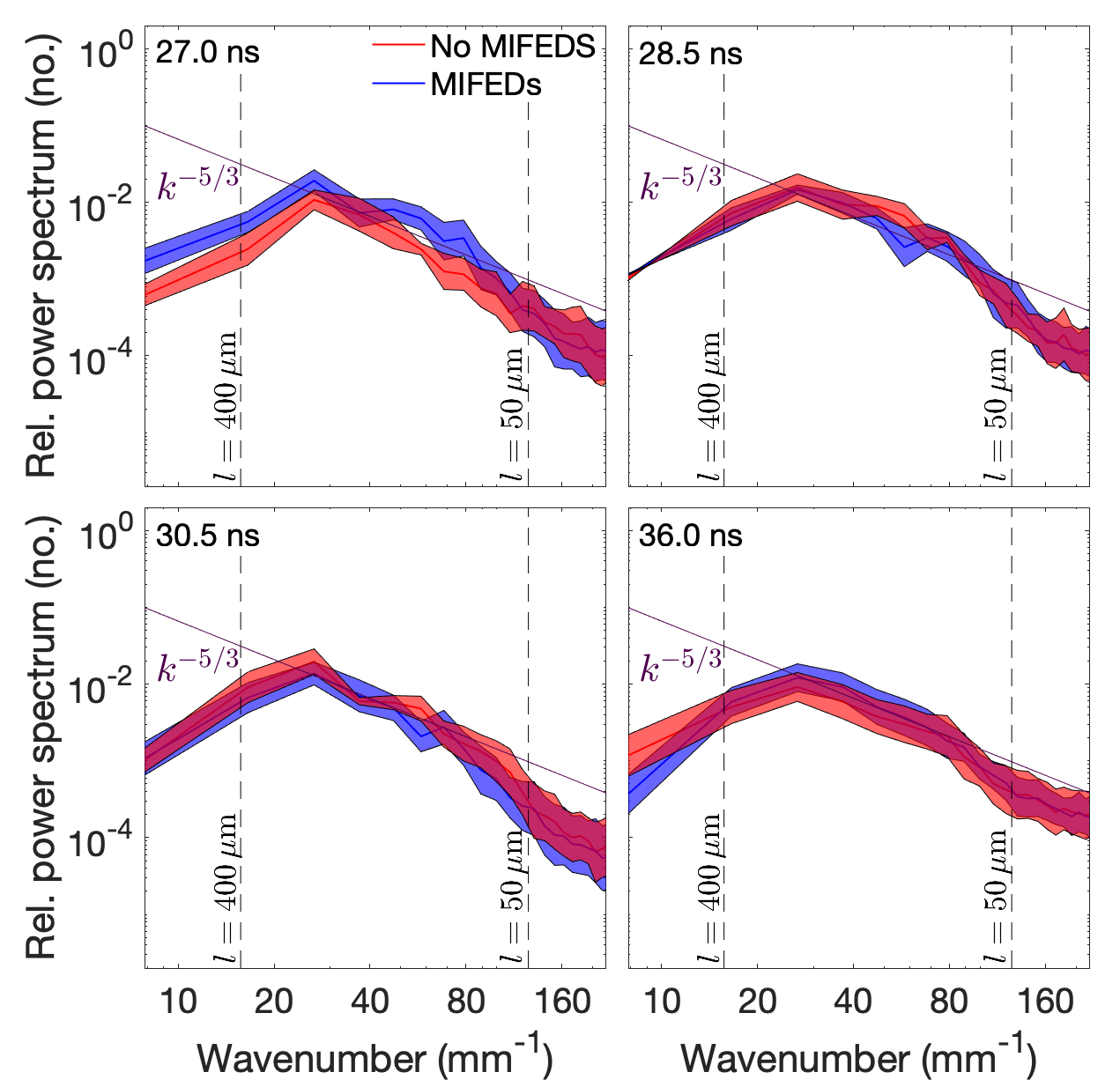}
\caption{\label{fig:Xrayimages_spectra} \emph{Spectral analysis of relative X-ray intensity fluctuations.} Inferred spectrum of density and velocity fluctuations in the plasma at four different times in the presence and absence of the MIFEDS magnetic field, respectively. We evaluate the (wavenumber-dependent) uncertainty of the measurement by combining the ${\sim} 10$\% uncertainty associated with the signal-to-noise ratio of the X-ray images with the standard error in the inferred spectrum that arises when the upper and lower regions of the interaction-region plasma are treated as independent samples.}
\end{figure}

Under the same assumptions of statistical homogeneity and isotropy, we can also determine the spectrum of turbulent density fluctuations in the plasma. Since the turbulent motions are subsonic, density behaves as a passive scalar, and thus its spectrum is simply that of the turbulent velocity field~\cite{Zhuravleva2014}. These spectra, which are shown Figure \ref{fig:Xrayimages_spectra}, have a similar shape irrespective of both the time of the measurement and whether the MIFEDS magnetic field is present or not: the spectral peak is at a wavenumber ${\sim} 2\pi/L \approx 20$ mm$^{-1}$, with spectra consistent with a Kolmogorov $-5/3$ power law at larger wavenumbers.

In summary, we conclude that while there are some modest differences in the plasma's initial dynamical evolution when the MIFEDS is present, the key properties of the plasma turbulence in the interaction region are essentially unaffected by it. 

\subsection{Diagnosing the plasma's physical state: Thomson scattering} \label{sec:TS}

\begin{figure*}[htbp]
\includegraphics[width=\linewidth]{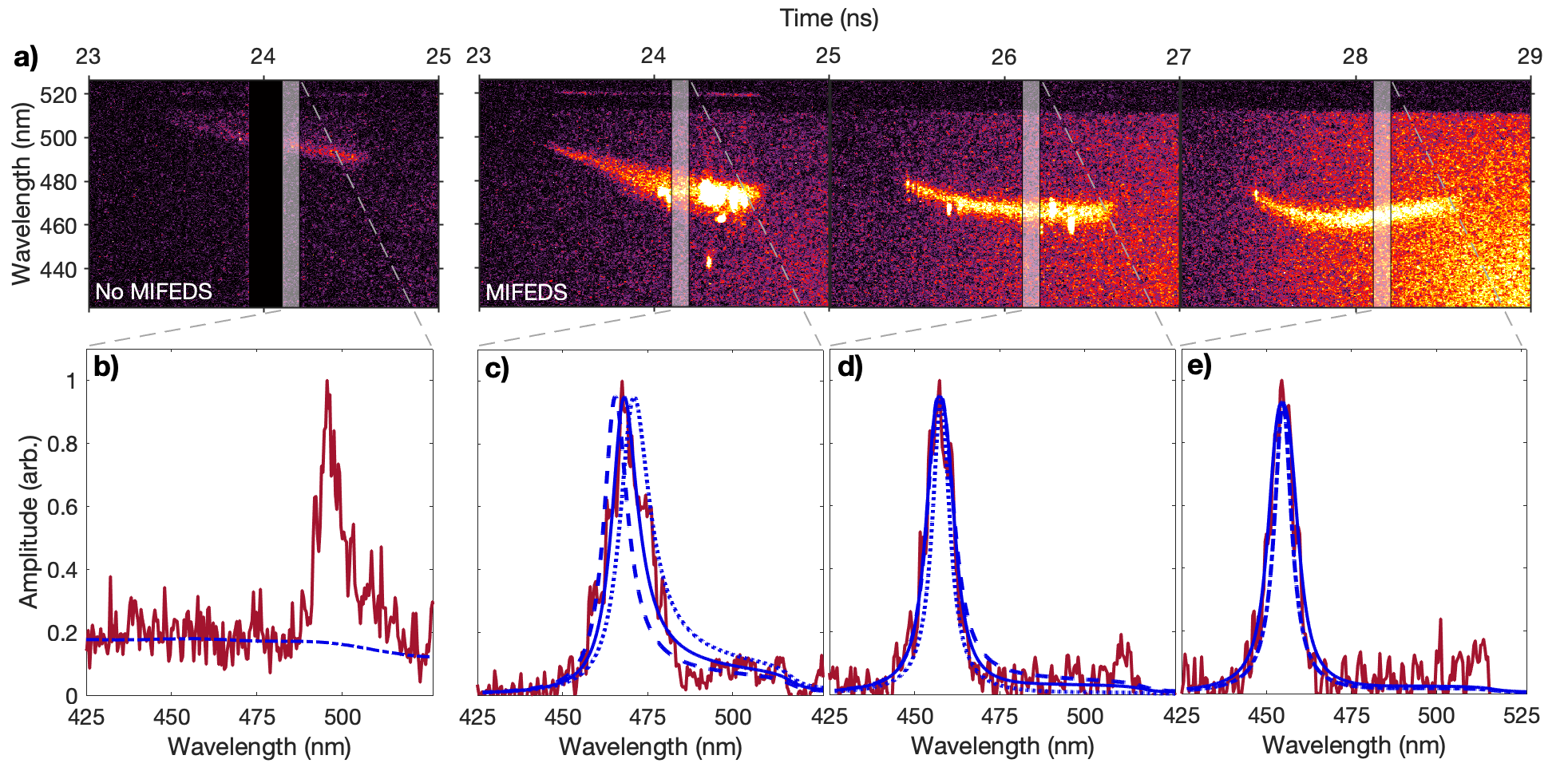}
\caption{\label{fig:TS_fits} \emph{Thomson-scattering data and fitting.} \textbf{a)} Time-resolved EPW spectral features obtained in the experiment. Data were successfully collected for one shot without MIFEDS and three shots with MIFEDS. For the no-MIFEDS experiment (far-left panel), the spectrometer used to detect the EPW spectral feature gave an erroneous output for a 200 ps interval centred at 24.0 ns; this output has been masked in the panel. The absolute magnitude of the signals is normalised to the same value in each image. \textbf{b)} 
Plot of the experimental signal (solid red line) obtained from the raw data shown in the far-left panel of a) by averaging over an 100 ps interval centred at 24.15 ns (viz., the interval indicated by the white translucent region). The blue dot-dashed line indicates the fit to the background signal that we subtract prior to constructing a best-fit model to the EPW spectral feature. \textbf{c)} Plot of the experimental signal (with the background subtracted) obtained from the raw data shown in the mid-left panel of a) by averaging over an 100 ps interval centred at 24.15 ns, along with three possible spectral fits with different mean electron number densities: one with $\bar{n}_e = n_{e,\mathrm{fit}} =  4.3\times10^{19} \, \mathrm{cm^{-3}}$ (solid blue line), a second with $\bar{n}_e = 0.85 n_{e,\mathrm{fit}} =  3.6\times10^{19} \, \mathrm{cm^{-3}}$ (dotted blue line), and a third with $\bar{n}_e = 1.15 n_{e,\mathrm{fit}} =  3.6\times10^{19} \, \mathrm{cm^{-3}}$ (dashed blue line). In all three cases, the assumed electron temperature is $T_e = 550$ eV. \textbf{d)} Same as c), but at  26.15 ns, along with three possible spectral fits with different electron temperatures (and mean electron number densities): one with $\bar{n}_e = n_{e,\mathrm{fit}} =  7.9 \times10^{19} \, \mathrm{cm^{-3}}$ and $T_e = T_{e,\mathrm{fit}} = 380$ eV (solid blue line), one with $\bar{n}_e = 1.1 n_{e,\mathrm{fit}} =  8.7 \times10^{19} \, \mathrm{cm^{-3}}$ and $T_e = 0.5 T_{e,\mathrm{fit}} = 190$ eV (dotted blue line), and one with $\bar{n}_e = 0.9 n_{e,\mathrm{fit}} =  7.1 \times10^{19} \, \mathrm{cm^{-3}}$ and $T_e = 1.5 T_{e,\mathrm{fit}} = 570$ eV (dashed blue line). In all cases, a Gaussian spread of densities with $\Delta n_e/\bar{n}_e = 0.25$ was assumed. \textbf{e)} Same as c), but at 28.15 ns, and with two possible spectral fits (both with $\bar{n}_e = n_{e,\mathrm{fit}} =  8.8 \times10^{19} \, \mathrm{cm^{-3}}$ and $T_e = T_{e,\mathrm{fit}} = 380$ eV): one with $\Delta n_e/\bar{n}_e = 0.25$ (solid blue line), and another with $\Delta n_e/\bar{n}_e = 0$ (dot-dashed blue line).}
\end{figure*}

{\color{black} For the Thomson-scattering diagnostic employed in the experiment, a 30 J, green laser-probe beam (with wavelength 526.5 nm) was focused onto the centre of the target (and hence the centre of the interaction-region plasma), and scattered light collected at an angle of $63^{\circ}$.} The orientation of the beam is shown in Figure \ref{fig:expdesign}. In this experiment, rather than carrying out measurements that were time-integrated over a 1 ns interval but spatially resolved along a 1.5 mm $\times$ 50 µm$^{2}$ cylindrical volume as was done previously~\cite{Bott2021}, we instead performed time-resolved measurements in a 50 µm$^{3}$ volume over the 1 ns interval. To obtain time-resolved measurements over the complete evolution of the interaction-region plasma, we repeated the experiment but applied the Thomson-scattering probe beam at different times. For a selection of different times around (and after) the formation of the interaction-region plasma, the `high-frequency' electron-plasma-wave (EPW) feature was successfully measured on a spectrometer; this data are shown in Figure \ref{fig:TS_fits}a. 
For reasons that remain uncertain, we were unable to detect successfully the low-frequency ion-acoustic-wave (IAW) feature; an anomalous signal saturated the spectrometer on which we had planned to detect this feature at the wavelengths over which it is typically observed.

We model the EPW feature using the well established theory of Thomson-scattering spectra that arise in plasmas~\cite{Evans1969}. In general, the Thomson-scattered spectrum $I({\bf k},\omega)$ at frequency $\omega$ of a laser probe beam with scattering wavevector ${\bf k}$ is given by 
\begin{equation}
I({\bf k},\omega) = N_e I_{0} \sigma_\mathrm{T} S({\bf k},\omega) \, ,
\end{equation}
where $N_e$ is the total number of electrons in the scattering volume, $I_0$ is the intensity of the incident laser probe, $\sigma_\mathrm{T}$ is the Thomson cross section for the scattering of a free electron, and $S({\bf k},\omega)$ is the dynamic form factor. We then adopt the Salpeter approximation for the form factor~\cite{Evans1969}, valid in a plasma with Maxwellian electron and ion distribution functions whose electron and ion temperatures $T_e$ and $T_i$ and electron and ion number densities $n_e$ and $n_i = n_e/Z$ (for $Z$ the ion charge) are such that $\alpha \equiv 1/k \lambda_{\rm D} \gg 1$, where $\lambda_D$ is the Debye length. This is a reasonable assumption for the experimental conditions. In the Salpeter approximation, 
\begin{equation}
  S({\bf k},\omega) \approx \frac{1}{k v_{\mathrm{th}e}} \Gamma_\alpha\!\left(\frac{\omega-\omega_0}{k v_{\mathrm{th}e}}\right)  
  \label{Salpeterapprox}
\end{equation}
at `high' frequencies $\omega -\omega_0 \sim k v_{\mathrm{th}e}$, where $v_{\mathrm{th}e}$ is the thermal electron speed, $\omega_0$ is the frequency of the incident laser probe,  
\begin{equation}
\Gamma_\alpha(x) \equiv \frac{\exp{\left(-x^2\right)}}{\sqrt{\pi} \left|1+\alpha^2 [1 + x Z(x)]\right|^2} \ ,  
\end{equation}
and $Z(x)$ is the plasma dispersion function~\cite{Fried1961}. 
It follows that the shape of the EPW spectral feature is directly related to $n_e$ and $T_e$ in a homogeneous plasma. Finally, in a turbulent plasma, the presence of density fluctuations in the interaction-region plasma typically gives rise to a range of electron number densities in the scattering volume. To capture this effect, we therefore assume that $n_e$ is isotropic and normally distributed in the scattering volume, with mean value $\bar{n}_e$ and standard deviation $\Delta n_e$. The 
EPW feature can then be modelled by
\begin{eqnarray}
  S_{\mathrm{EPW}}({\bf k},\omega) & \approx & \frac{1}{\sqrt{\pi} \Delta n_e} \int \mathrm{d} \tilde{n}_{e} \, \exp{\left[-\frac{\left(\tilde{n}_{e}-\bar{n}_{e}\right)^2}{\Delta n_e^2}\right]} \nonumber\\
  && \qquad \times \frac{1}{k v_{\mathrm{th}e}} \Gamma_\alpha\!\left(\frac{\omega-\omega_0}{k v_{\mathrm{th}e}}\right). 
  \label{Salpeterapprox_elec}
\end{eqnarray}
Qualitatively, for frequencies $\omega > \omega_0$, this feature has a single peak. The peak's position is sensitive to $\bar{n}_e$ and, to a much lesser extent, to $T_e$, while its width is sensitive to $T_e$ and $\Delta n_e$. 

Having established a model for the EPW feature, we fit the data as follows. First, we perform a background subtraction in order to remove signals on the spectrometer that are unrelated to the EPW feature. The background signal is approximated using (Gaussian-filtered) samples taken just before and after the duration of the Thomson-scattering probe beam and then interpolating those signals to a given time (a typical background signal determined using this approach is shown in Figure \ref{fig:TS_fits}b). Then, we fit equation~(\ref{Salpeterapprox_elec}) for particular choices of $\bar{n}_e$, $T_e$ and $\Delta n_e$ against 100-ps averaged samples of data, substituting for $\omega$ in terms of the wavelength $\lambda$ using the dispersion relation $\omega = \omega(\lambda)$ of a light wave passing through a plasma. The approach of choosing $\bar{n}_e$, $T_e$ and $\Delta n_e$ differs depending on whether we are fitting EPW features close to the collision of the plasma jets, or subsequent to it. In the former case, we assume that turbulence has not yet developed and set $\Delta n_e = 0$; we then vary $\bar{n}_e$ and $T_e$ to obtain the best fit of the peak's position and width. In the latter case, we are faced with a degeneracy, with both changes in $\Delta n_e$ and $T_e$ having very similar effects on the width of the spectral peak. To overcome this degeneracy, we infer an estimate for $\Delta n_e$ from the measurements of relative density fluctuations obtained using the X-ray imaging diagnostic (see section \ref{sec:Xray}): namely, we assume that the rms of density fluctuations on the scale $l_{\rm T}$ of the Thomson-scattering volume is related to the rms of density fluctuations at the turbulence's integral scale via a Kolomogorov scaling $\Delta n_e/\bar{n}_e \approx (\Delta n_e/\bar{n}_e)_{l_n} (l_{\rm T}/l_n)^{1/3} \approx 0.25$. {\color{black}The validity of this assumption has been tested by our previous experiments~\citep{Bott2021}, in which explicit measurements of $\Delta n_e/\bar{n}_e$ were made (possible due to successful simultaneous measurements of both the IAW and EPW features); these measurements indeed recovered similar values to those inferred from the X-ray images.} We then once again adjust $\bar{n}_e$ and $T_e$ to give the best fit of the EPW spectral feature's position and width. 

\begin{figure}[htbp]
\includegraphics[width=\linewidth]{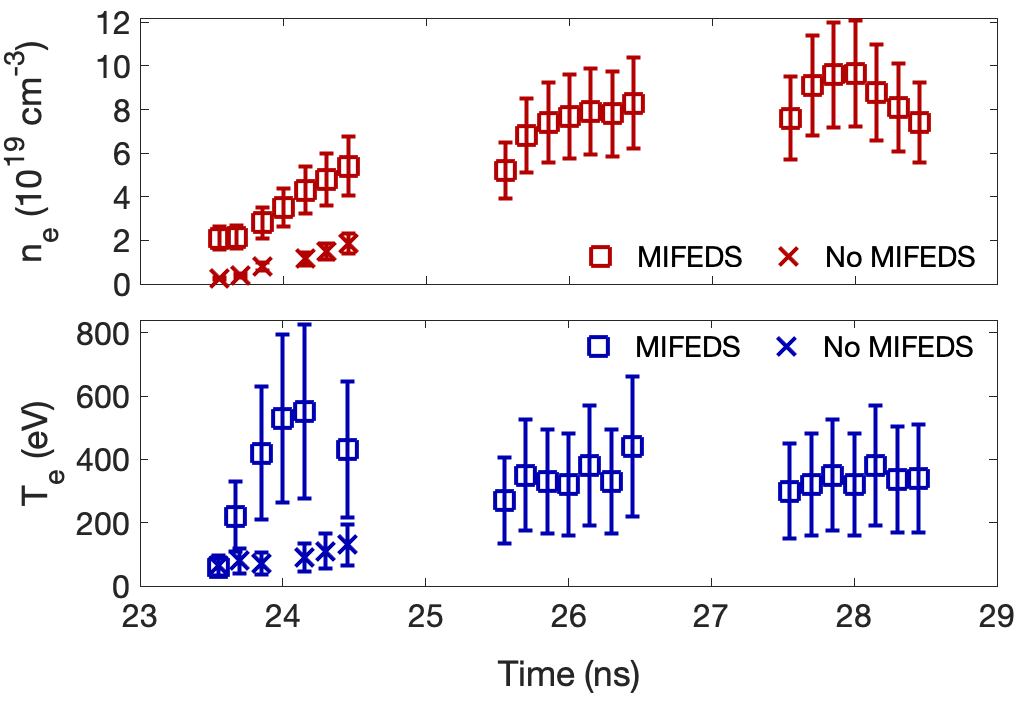}
\caption{\label{fig:TS_outputs} \emph{Thomson-scattering derived measurements of the plasma's physical state.} Evolution of the electron number density and temperature in the presence and absence of the MIFEDS magnetic field around and just after collision, as inferred from spectral fits. The uncertainties are determined from the sensitivity of the fits: $\pm 25\%$ for $n_e$, and $\pm 50\%$ for $T_e$. In the MIFEDS experiment, we were unable to construct a fit for the electron temperature at 24.3 ns on account of distortion of the EPW signal. }
\end{figure}

Once a best fit is obtained, we assess the sensitivity of the fits by first determining how the peak's position responds to changes in $\bar{n}_e$ while keeping $T_e$ fixed (see Figure \ref{fig:TS_fits}c); next, we vary $T_e$ and $\bar{n}_e$ concurrently in such a way that the peak position remains fixed, but its width changes (see Figure \ref{fig:TS_fits}d). We conclude from this analysis that the combined sensitivity of the fits to changes in $n_e$ is $\pm 25\%$, while the sensitivity to changes in $T_e$ is $\pm 50\%$ (taking correlated uncertainties into account). Finally, the sensitivity of the fits to our assumptions concerning the magnitude of $\Delta n_e$ is illustrated in Figure \ref{fig:TS_fits}e; we find that, in the absence of any turbulent broadening, inferred electron temperatures would be ${\sim}50\%$ larger.  The mean electron number densities $\bar{n}_e$ and temperatures $T_e$ derived from the fitting procedure for all of the data are shown in Figure \ref{fig:TS_outputs}.

For the time interval of 23.5 ns to 24.5 ns during which we have data for both the MIFEDS and no MIFEDS experiments, we observe significant differences in the physical properties of the plasma: namely, the inferred values of $n_e$ are much larger in the former case, and rapid heating of the electrons is observed in the presence of MIFEDS, but not in its absence. A compelling explanation for these observations is that the collision between the opposing plasma jets occurs ${\sim}$1 ns earlier -- at ${\sim}24$ ns -- in the MIFEDS experiments than in the no-MIFEDS experiments (in which collision occurs at ${\sim}$25 ns based on prior measurements~\cite{Bott2021}). The physical origin of this timing difference can be attributed to a dynamical collimation effect of the MIFEDS magnetic field on the jets. Using the data in Figure \ref{fig:TS_outputs} to quantify the parameters of the jets just before they collide ($\rho_{\rm jet}\approx 6\times 10^{-5} \, \mathrm{g}/\mathrm{cm}^3, T_{e\mathrm{,jet}}\approx T_{i\mathrm{,jet}}\approx 100 \; \mathrm{eV}$), it follows that the characteristic kinetic-energy density of the jets’ transverse expansion (which we estimate as $\epsilon_{\rm K,jet\perp} = \rho_{\rm jet} u_{\rm jet\perp}^2/2\approx 2.9\times 10^{8}$ erg cm$^{-3}$ by assuming that the expansion velocity $u_{\rm jet\perp}$ is given by the sound speed $c_{\rm s}\approx 1.0\times10^7$  cm/s in the jet) is comparable to the magnetic-energy density $\epsilon_{B_0} = B_0^2/8\pi \approx 2.5\times 10^{8}$ erg cm$^{-3}$ of the MIFEDS magnetic field. Therefore, the transverse expansion of the jets is at least partially inhibited by the MIFEDS magnetic field, a conclusion that is supported by the X-ray imaging observations (see Figure \ref{fig:Xrayimages}, top row).
It is in turn plausible that this collimation is associated with a small increase in the jets' parallel velocity; the inferred collision timing difference is consistent with an ${\sim}$5\% increase in the initial jet velocities over the no-MIFEDS experiments to $u_{\rm jet}\approx 2.4 \times 10^7$ cm/s. We note that, although the MIFEDS magnetic field does seem to have a dynamical effect on the plasma jets, the total characteristic kinetic-energy density of either jet ($\epsilon_{\rm K,jet} = \rho_{\rm jet} u_{\rm jet}^2/2\approx 1.7\times 10^{9}$ erg) is indeed significantly larger than $\epsilon_{B_0}$, as claimed in section \ref{sec:exp_design}. 

By contrast, the Thomson-scattering measurements of the interaction-region plasma's parameters in the MIFEDS experiments post collision are similar to those of no-MIFEDS experiments. A few ns after collision, we obtain characteristic temperatures $T_e \approx T_i\approx250$-$450$ eV, and electron number densities $n_e\approx(0.6$-$1.0)\times10^{20} \, \mathrm{cm^{-3}}$, parameters that are close to those inferred from previous experimental data collected on the Omega Laser Facility~\cite{Tzeferacos2018,Bott2021}. While we do not have a direct measurement of the rms turbulent velocity in the MIFEDS experiments, the inferred ${\sim}$5\% difference in the incident jet velocities between the MIFEDS and no-MIFEDS experiments is small enough that, given the much larger ${\sim} 40\%$ uncertainty of the turbulent-velocity measurements in previous OMEGA experiments, we believe it to be reasonable to infer that the turbulent velocities in the no-MIFEDS and MIFEDS cases are similar ($u_{\rm rms} \approx 110$ km/s). Therefore, we conclude that a turbulent plasma with a large magnetic Reynolds number ($\mathrm{Rm} \approx 200$-$450$) is indeed realized in this experiment, with the plasma's turbulent dynamics being affected minimally by the MIFEDS magnetic field. This latter conclusion is consistent with that derived from the X-ray imaging diagnostic.

\subsection{Diagnosing the plasma's magnetic fields: proton radiography} \label{sec:PRAD}

{\color{black} The source of the protons for the proton-radiography diagnostic utilised in the experiment was a spherical aluminium-coated Si$\mathrm{O}_2$ capsule (thickness 2 µm, diameter 420 µm) filled with 18 atm D$^{3}$He gas, with the centre of the capsule located at a distance $r_{\rm s} = 1$ cm away from the target's centre.} Upon irradiation with ${\sim}$8 kJ of laser energy over a 1 ns interval, the capsule implodes on a timescale of a few hundred picoseconds; nuclear fusion reactions then generate ${\sim} 10^9$ 3.3 MeV and 15.0 MeV protons, which subsequently stream away from the capsule in all directions. A fraction of these pass through the interaction-region plasma, before reaching a detector (located a distance $r_{\rm d} = 27$ cm away from the target's centre) consisting of layers of the nuclear track detector CR-39 and metallic filters. The detector images the 3.3 and 15.0 MeV protons independently. 
Both the proton source and the detector have been carefully characterised in numerous prior studies~\cite{Seguin2003,Li2007,Manuel2012}. 
In contrast to previous Omega experiments investigating turbulent-dynamo processes, proton radiography in this experiment was performed in a side-on configuration with respect to the interaction-region plasma, in order to accommodate the change in orientation of the XRFC diagnostic. To obtain radiography measurements at different times, we repeated the experiments but changed the relative timing of the capsule implosion with respect to the drive beams incident on the CH foils.  

In our experiment, proton-radiography data provide a wealth of information about the magnetic fields encountered by the protons as they travel from the source to the detector. In the absence of any such fields, the proton-radiography beam would retain its inherent homogeneity, and thus the proton flux measured at the detector would be close to uniform. In reality, magnetic fields are encountered, and the action of Lorentz forces associated with these fields causes small deflections in the protons' trajectories, changing the location at which they arrive at the detector. In general laser-plasma experiments, electric fields could also cause these deflections; however, for laser-plasma dynamo experiments such as ours, their impact is minimal~\cite{Tzeferacos2018}. If the proton beam is partially blocked prior to its interaction with the magnetic fields, deflections of the beam can be directly visualised, providing a very simple way to assess the path-integrated magnetic field. 
If the magnetic fields are also spatially heterogeneous, this can lead to significant transverse inhomogeneities in the proton beam as seen at the detector. Such inhomgeneities can be analysed quantitatively using a (now well established~\cite{Tzeferacos2018,Bott2021,Bott2021b,Palmer2019,Schaeffer2020,Campbell2020,Tubman2021}) technique that directly relates proton-flux inhomogeneities to the magnetic field path-integrated along the trajectory of the beam protons using a field-reconstruction algorithm~\cite{Bott2017,Kasim2019}; the technique is formally valid under a set of assumptions that are satisfied in the proton radiography set-up, and has been cross-validated on the Omega Laser Facility using Faraday rotation measurements~\cite{Rigby2018}. 

\begin{figure}[htbp]
\includegraphics[width=\linewidth]{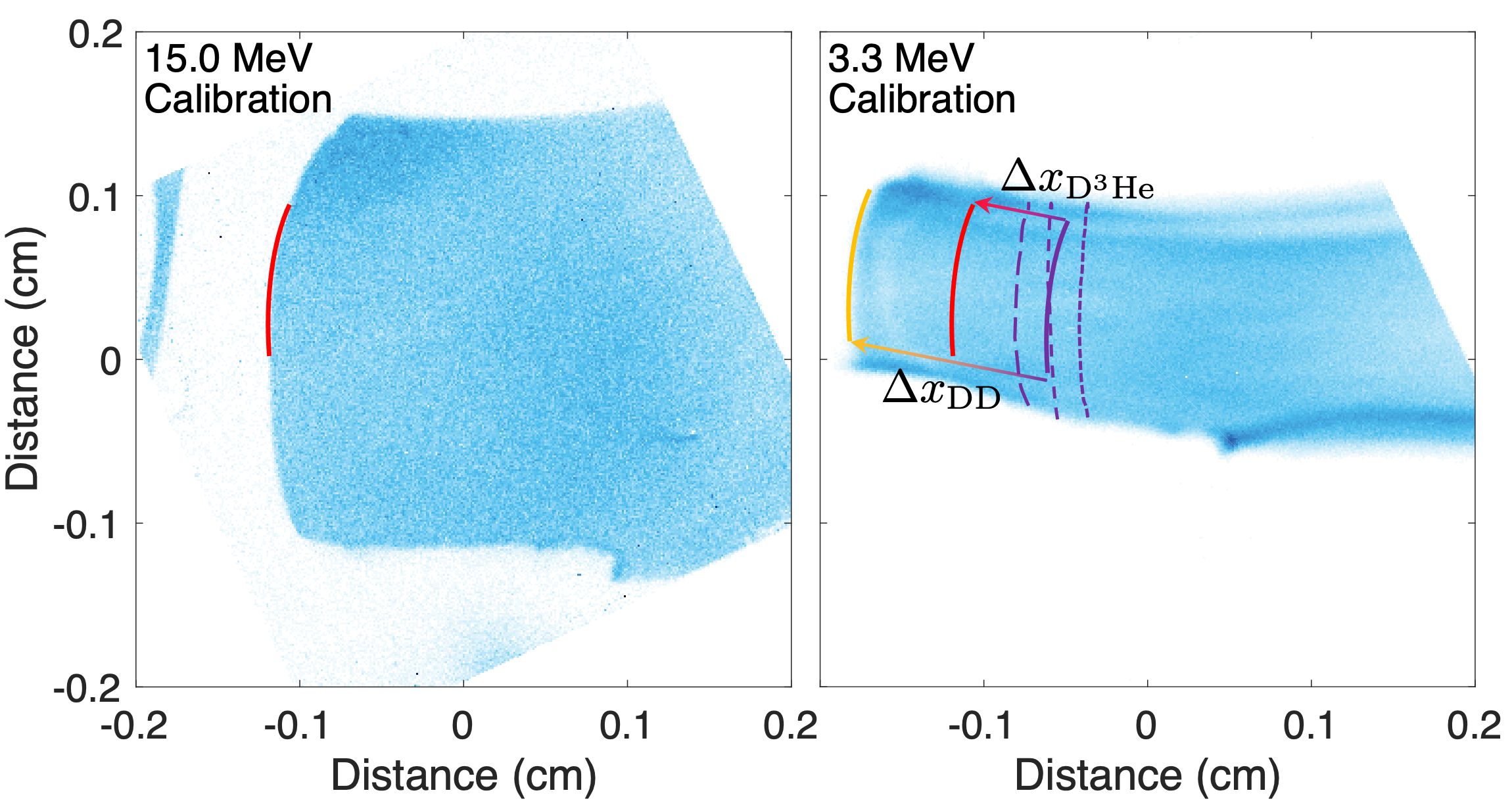}
\caption{\label{fig:PRADcal} \emph{Calibration measurement of the MIFEDS magnetic field with proton radiography.} Left panel: 15.0-MeV proton radiograph of the target in the absence of any drive beams, but with MIFEDS on. The axes of the image, which has a $\times$28 magnification, are rescaled so that lengths are directly comparable with the plasma's scale. The reported pixel counts are normalised to their mean value (${\sim}$60 protons/pixel) in a 0.1 cm by 0.1 cm square whose midpoint is at the centre of each image. Right panel: 3.3-MeV proton radiograph of same set-up. In both panels, the red line marks the apparent boundary of the 15.0 MeV proton beam, while the gold line marks the apparent boundary of the 3.3 MeV proton beam. The solid purple line marks the boundary of both proton beams in the absence of any magnetic fields that is inferred from the relative displacement of the apparent boundary of the 15.0 MeV and 3.3 MeV beams; the short-dashed, mid-dashed, and long-dashed lines denote the observed boundary of the 15.0 MeV proton beams at 25.2 ns, 31.2 and 38.7 ns, respectively, in the no-MIFEDS experiments. In these images, the line of centers is vertical and the targets and grids lie at the top and the bottom of it.}
\end{figure}

The proton-radiography diagnostic was first used to perform a calibration measurement of the MIFEDS-generated magnetic field, confirming that it has the expected strength and orientation. For this measurement, the MIFEDS was activated and the D$^{3}$He capsule imploded, but the drive beams incident on the target's CH foils were not fired. The resulting 15.0-MeV and 3.3-MeV proton radiographs are shown in Figure \ref{fig:PRADcal}. 
For a magnetic field that is oriented as indicated in Figure \ref{fig:expdesign} (viz., approximately parallel to the line of centres), it is to be expected that the protons that pass through the centre of the target assembly would be displaced towards the left side of the detector, with the 3.3 MeV protons displaced further than 15.0 MeV protons. This is indeed what is observed in Figure \ref{fig:PRADcal}: namely, before passing through the centre of the target assembly, part of both proton beams is blocked by a wire associated with the MIFEDS, and the apparent boundary of this wire is further on the left in the 3.3-MeV proton radiograph than in the 15.0-MeV radiograph. 

More quantitatively, the path-integrated magnetic field experienced by protons traversing the MIFEDS magnetic field can be explicitly estimated from the relative displacement of the boundary. In a point-projection radiography set-up, it can be shown~\cite{Kugland2012} that the displacements  $\Delta x_{\rm D^{3} He}$ and $\Delta x_{\rm DD}$ of protons from their undeflected position on the detector are given by $\Delta x_{\rm D^{3} He} \approx r_{\rm d} \Delta v_{\rm D^{3} He}/v_{\rm D^{3} He}$ and $\Delta x_{\rm DD} \approx r_{\rm d} \Delta v_{\rm DD}/v_{\rm DD}$, respectively (where $\Delta v_{\rm D^{3} He}$ and $\Delta v_{\rm DD}$ are the velocity perturbations of 15.0 MeV and 3.3 MeV protons acquired due to the interaction with the magnetic field, and $v_{\rm D^{3} He}$ and $v_{\rm DD}$ are the initial speeds of the 15.0 MeV and 3.3 MeV protons). In the limit of small deflections, $\Delta v_{\rm D^{3} He} \approx \Delta v_{\rm DD} \approx e \int B_{\perp} \mathrm{d}s/m_p c$ is independent of the proton velocity (where $B_{\perp}$ is the component of the magnetic field perpendicular to the direction of the proton beam, $e$ is the elementary charge, $c$ the speed of light, and $m_p$ the proton mass), and so it follows that
\begin{equation}
\int B_{\perp} \mathrm{d}s \approx \frac{m_p c v_{\rm D^{3} He} v_{\rm DD}}{e (v_{\rm D^{3} He}-v_{\rm DD})} \frac{\Delta x_{\rm DD}-\Delta x_{\rm D^{3} He}}{r_{\rm d}} \, . \label{pathB_MIFEDScal_est}
\end{equation}
We find that $\Delta x_{\rm DD}-\Delta x_{\rm D^{3} He} \approx 1.7$ cm; equation~(\ref{pathB_MIFEDScal_est}) then gives $\int B_{\perp} \mathrm{d}s \approx 25$ kG cm. This is consistent with theoretical expectations of the MIFEDS magnetic field, for which $B_{\perp} \approx 80$ kG across a region of extent $l_{\rm path} \approx 0.3$ cm. As a sanity check of the validity of this approach, in the right panel of Figure \ref{fig:PRADcal} we compare the position of the proton beam's undeflected boundary inferred from our calculation of $\int B_{\perp} \mathrm{d}s$ with direct measurements of this quantity in no-MIFEDS experiments (in which it is anticipated that the boundary of the proton beam is unperturbed). We find reasonable agreement, given the uncertainties arising from the positioning of the MIFEDS wire due to inconsistent target fabrication. 

\begin{figure}[htbp]
\includegraphics[width=\linewidth]{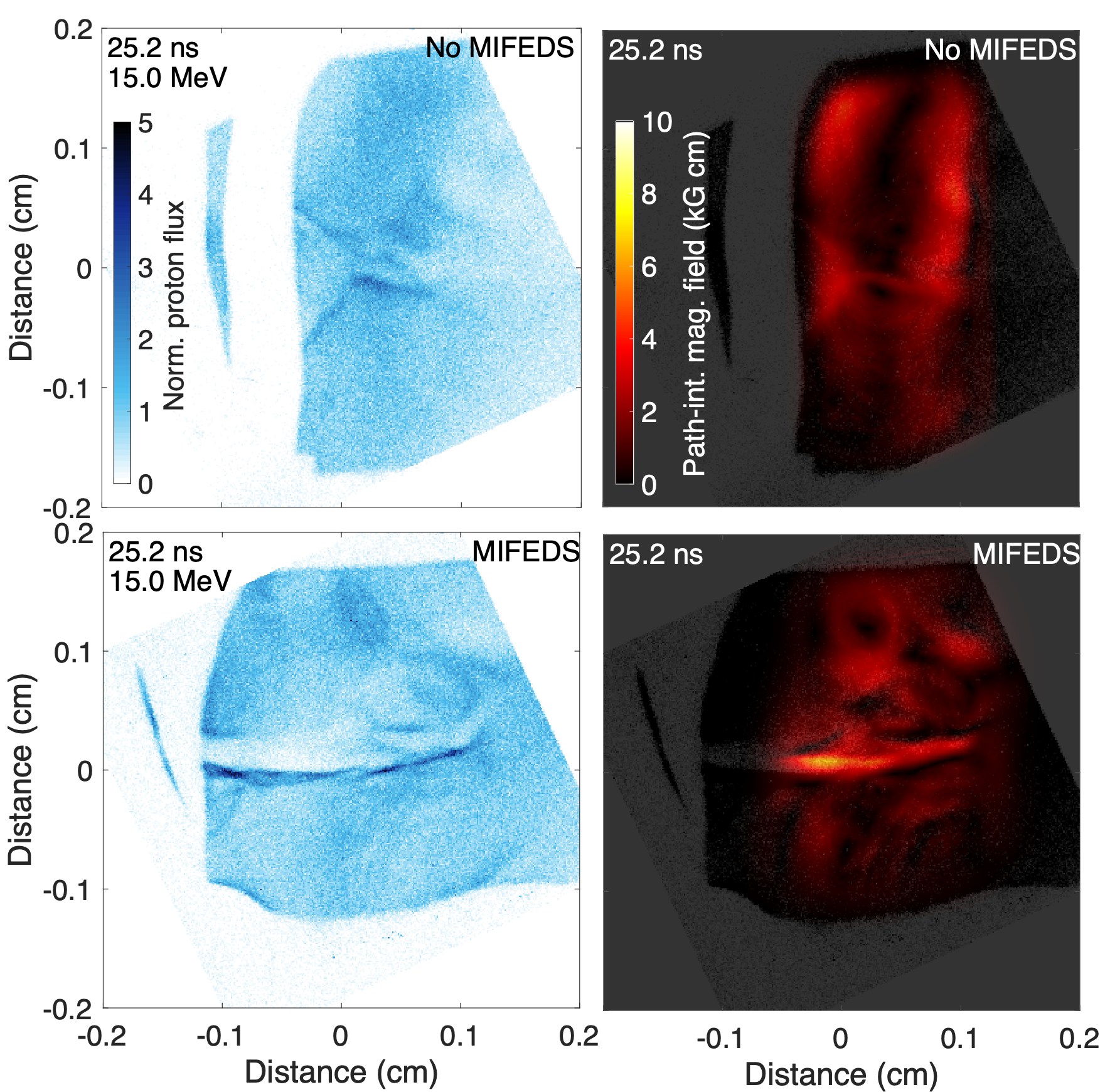}
\caption{\label{fig:PRAD1} \emph{Measurements of magnetic fields at collision with proton radiography.} Left column: 15.0-MeV proton radiographs in the presence/absence of the MIFEDS at 25.2 ns (at a time close to the collision of the plasma jets). The pixel counts of each image are normalised to their mean value (${\sim}$60 protons/pixel) in a 0.1 cm by 0.1 cm square whose midpoint is at the centre of each image. In these images, the line of centers is vertical and the targets and grids lie at the top and the bottom of it. The location of the interaction region is offset by ${\sim}$0.05 cm leftwards in the MIFEDS image due to the effect of the large-scale MIFEDS magnetic field. Right column: magnitude of the `small-scale’ components of the path-integrated magnetic field that is perpendicular to the trajectory of the proton radiography beam. In each case, we determine this quantity (using a field-reconstruction algorithm – see main text) over a region that is approximately coincident with the location of the interaction-region plasma, and only show those fluctuations in the path-integrated magnetic field whose characteristic scale is smaller than the characteristic size of the region analysed. In the case when the MIFEDS is on, we recover a large-scale path-integrated magnetic field in addition to the small-scale path-integrated field that causes the deflection of protons leftwards that we discuss in the text. To enable direct comparison, this field is not shown, and the positioning of the small-scale path-integrated field in these cases is adjusted to take this deflection into account.}
\end{figure}

Having calibrated the MIFEDS magnetic field strength and morphology, we then performed comparative measurements of magnetic fields arising in the turbulent interaction-region plasma with and without the MIFEDS switched on. 15.0 MeV proton radiographs recorded just after collision are shown in Figure \ref{fig:PRAD1}, left column. 
It is clear that the inhomogeneities of proton flux are more pronounced in the MIFEDS experiments than in the no-MIFEDS ones; because these inhomogeneities can be attributed to deflection of the proton beam by Lorentz forces associated with non-uniform magnetic fields present in the plasma~\cite{Kugland2012}, this implies stronger seed fields. 
Two-dimensional maps of the path-integrated magnetic field reconstructed using a field-reconstruction algorithm~\cite{Bott2017} are shown in Figure \ref{fig:PRAD1}, right column; when the MIFEDS is on, we estimate that the initial magnetic-field strength in the interaction-region plasma is
\begin{equation}
B_0\approx 60 \left[\frac{\int B_{\perp} \mathrm{d}s}{6 \, \mathrm{kG} \, \mathrm{cm}}\right]\left[\frac{l_{\rm path}}{0.1 \, \mathrm{cm}}\right]^{-1} \, \mathrm{kG} ,
\end{equation}
(where $l_{\rm path}$ is the path-length of the protons through the interaction-region plasma). This value is comparable with the MIFEDS field in the absence of the plasma jets, and is also much larger than the Biermann battery-generated seed fields observed in no-MIFEDS experiments ($B_0\approx 10$ kG). We note that the observed structure of the seed field in the MIFEDS experiments at the time of collision is qualitatively distinct to the MIFEDS field in the absence of the interaction-region plasma. This is most plausibly explained by the interaction of the plasma jets with the MIFEDS field; the former's kinetic-energy density is approximately ten times greater than the magnetic-energy density of the MIFEDS magnetic fields, and the jets' magnetic Reynolds number is significantly larger than unity ($\mathrm{Rm}_{\rm jet} \approx 50$-$90$), which results in the MIFEDS magnetic field being advected with the plasma jets as they expand towards each other. 

\begin{figure}[htbp]
\includegraphics[width=\linewidth]{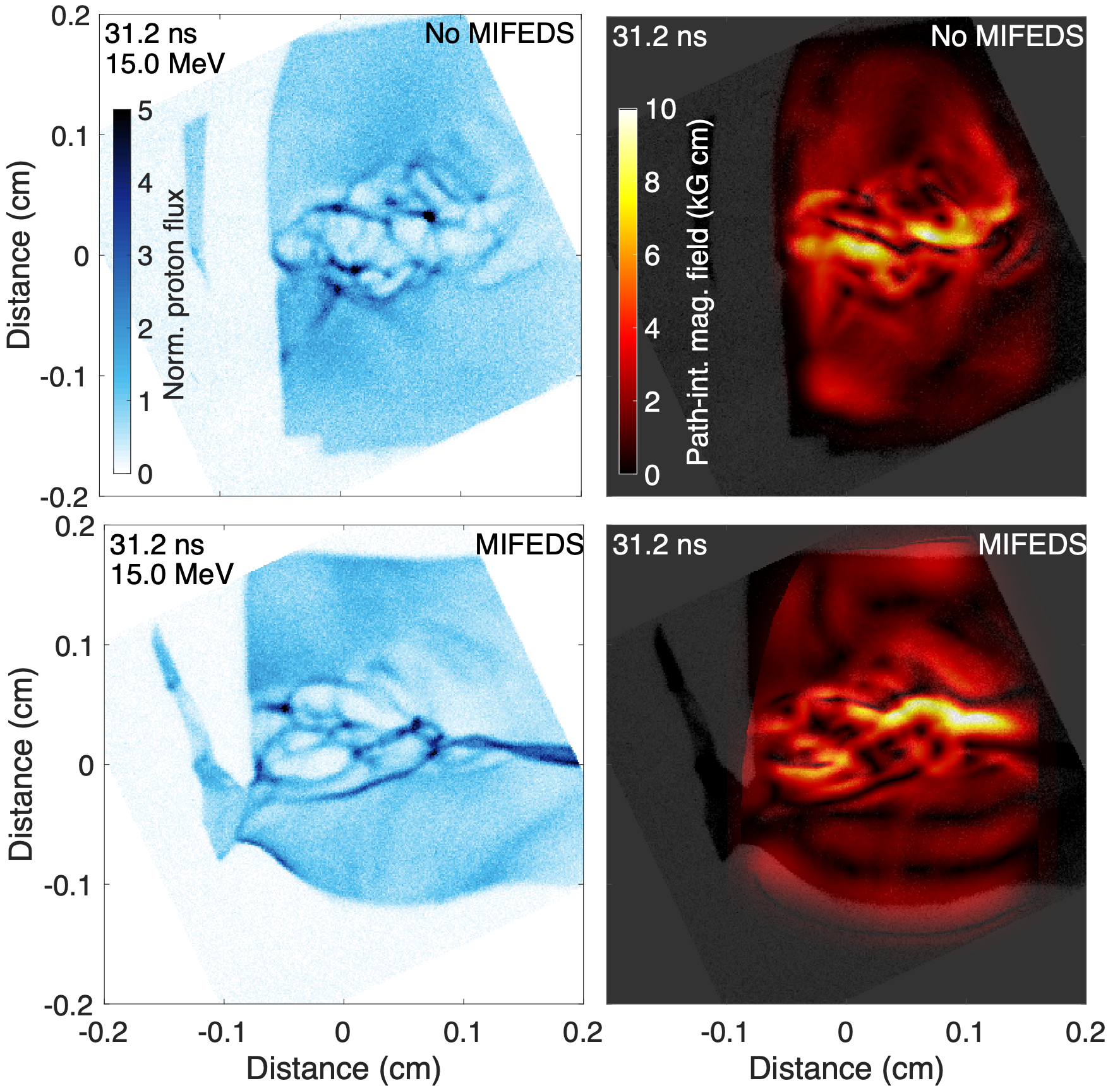}
\caption{\label{fig:PRAD2} \emph{Measurements of magnetic fields post collision with proton radiography.} Left column: 15.0-MeV proton radiographs in the presence/absence of the MIFEDS ${\sim}$6 ns after collision. Each image is normalised to its mean value (${\sim}$60 protons/pixel) in a 0.1 cm by 0.1 cm area in the centre of each image. In these images, the line of centers is vertical and the targets and grids lie at the top and the bottom of it. The location of the interaction region is offset by ${\sim}$0.05 cm leftwards in the MIFEDS image due to the effect of the large-scale MIFEDS magnetic field. The long horizontal feature in the MIFEDS image lies to the right of the interaction region (see text). Right column: magnitude of the small-scale components of the (perpendicular) path-integrated magnetic field.}
\end{figure}

\begin{figure}[htbp]
\includegraphics[width=\linewidth]{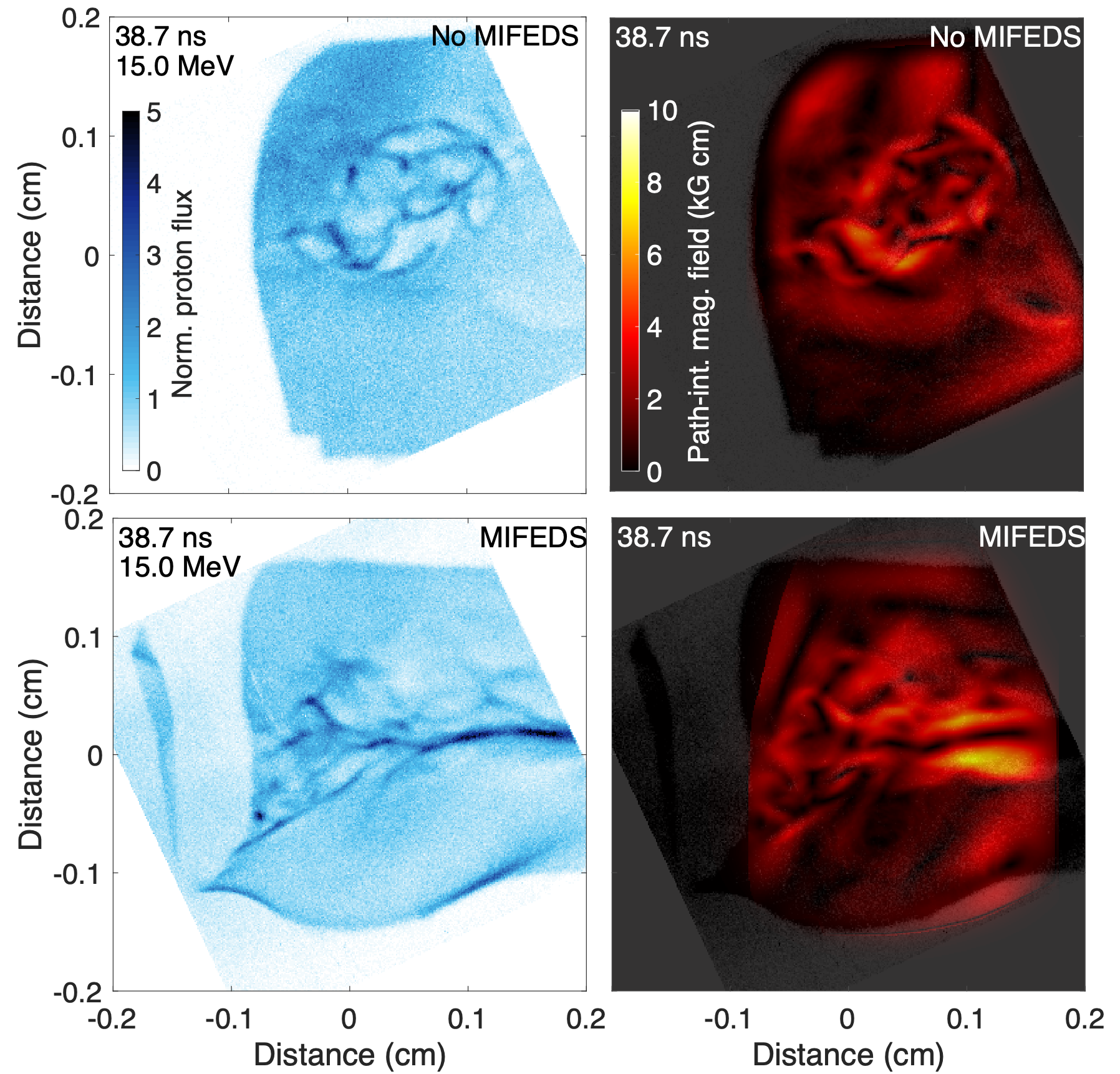}
\caption{\label{fig:PRAD3} \emph{Measurements of magnetic fields at late times with proton radiography.} Left column: 15.0-MeV proton radiographs in the presence/absence of the MIFEDS 13.5 ns after collision. The proton flux is detected using a CR-39 detector stack. The pixel counts of each image are normalised to their mean value (${\sim}$60 protons/pixel) in a 0.1 cm by 0.1 cm square whose midpoint is at the centre of each image. In these images, the line of centers is vertical and the targets and grids lie at the top and the bottom of it. The location of the interaction region is offset by ${\sim}$0.05 cm leftwards in the MIFEDS image due to the effect of the large-scale MIFEDS magnetic field. The long horizontal feature in the MIFEDS image lies to the right of the interaction region (see text). Right column: magnitude of the small-scale components of the (perpendicular) path-integrated magnetic field.}
\end{figure}

\begin{figure}[htbp]
\includegraphics[width=\linewidth]{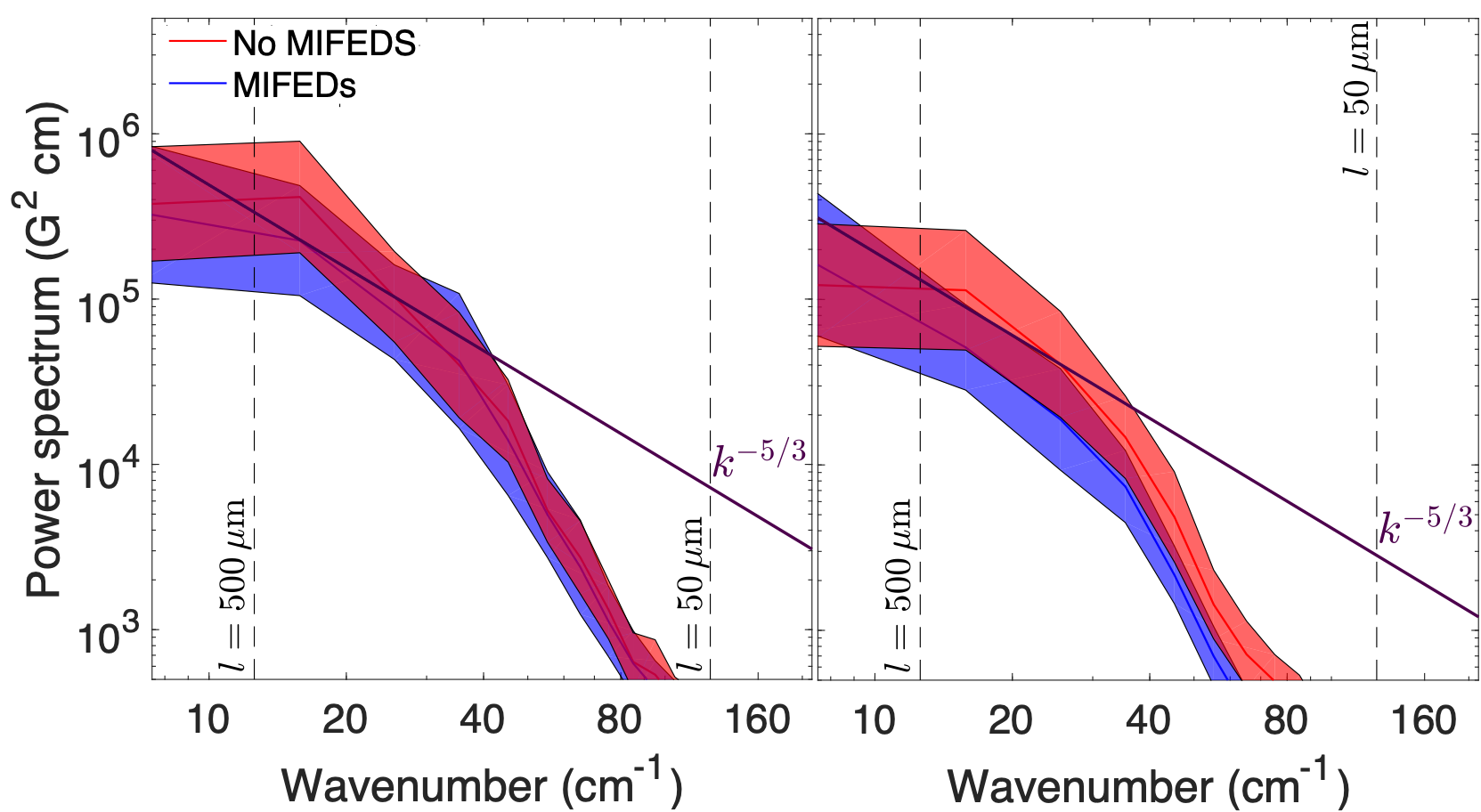}
\caption{\label{fig:PRAD4} \emph{Measurements of magnetic-energy spectra using proton radiography.} Magnetic-energy spectra inferred from proton radiography data under the assumption of homogeneous and isotropic stochastic magnetic fields. Spectra obtained from no-MIFEDS experiments are shown in red, and those from MIFEDS experiments in blue. The nominal limit on the resolution due to the finite size of the proton source is indicated on each plot; however, the actual resolution scale is observed to be a few times larger than indicated due to a systematic blurring of the proton-radiography data that stems from self-intersection of the proton beam prior to its detection (the self-intersection is caused by small-scale stochastic magnetic fields in the plasma~\cite{Bott2017,Bott2021b}). The uncertainty of the measurement of the spectrum is estimated by assuming that the interaction-region plasma is homogeneous, and then treating the interaction region's left- and right-hand sides as independent samples. Left panel: magnetic-energy spectra at 31.2 ns after collision. Right panel: magnetic-energy spectra at 38.7 ns.}
\end{figure}

In contrast to our findings close to the jet collision, both the (stochastic) proton-flux inhomogeneities and the reconstructed path-integrated magnetic fields are much more similar over one driving-scale turbulent eddy-turnover time (${\sim}$6 ns) after collision (see Figure \ref{fig:PRAD2}), and also over three driving-scale eddy-turnover times (${\sim}$13.5 ns) after the collision (see Figure \ref{fig:PRAD3}). 
Qualitatively, the proton radiographs from the MIFEDS and no-MIFEDS experiments are not completely identical: a significant proton-flux inhomogeneity with a magnitude much greater than the mean proton flux of the image, which is associated with the interaction of the MIFEDS field with the edge of the interaction-region plasma, is evident in the former on the right of the radiograph. However, the stochastic proton-flux inhomogeneities in the centre of the interaction-region plasma are much harder to distinguish, as are the stochastic path-integrated fields. Assuming that the magnetic field has isotropic and homogeneous statistics, we estimate the rms magnetic-field strength $B_{\rm rms}$ from the path-integrated magnetic-field maps via the relation $B_{\rm rms}\approx \int B_{\perp} \mathrm{d}s/\sqrt{\ell_B l_{\rm path}}$ (where $\ell_B$ is the field’s correlation length)~\cite{Bott2017}. For both the MIFEDS and no-MIFEDS experiments ${\sim}$6 ns after collision, we obtain 
\begin{equation}
B_{\rm rms} \approx 100 \left[\frac{\int B_{\perp} \mathrm{d}s}{4.5 \, \mathrm{kG} \, \mathrm{cm}}\right]\left[\frac{\ell_{B}}{0.01 \, \mathrm{cm}}\right]^{-1/2} \left[\frac{l_{\rm path}}{0.2 \, \mathrm{cm}}\right]^{-1/2} \, \mathrm{kG} \, .   
\end{equation}
This is comparable to the measured values of $B_{\rm rms}$ in previous experiments with similar Rm~\cite{Tzeferacos2018,Bott2021}. We can also estimate the magnetic-energy spectrum via the relation
\begin{equation}
E_B(k) = \frac{1}{4 \pi^2 l_{\rm path}}  k E_{\rm path}\!(k) \, , \label{Especform}
\end{equation}
where $E_{\rm path}\!(k)$ is the spectrum of the path-integrated magnetic fields; we note that the effective resolution of the proton-radiography diagnostic is ${\sim}100$-$200$ µm, so we only obtain the spectrum of the fields whose scale is comparable to the integral scale $l_{n}$ of the turbulence. The magnetic-energy spectra for both MIFEDS and no-MIFEDS experiments at 31.2 ns are shown in Figure \ref{fig:PRAD4}, left panel; within the uncertainty of the measurement, they are the same. 
The similarity of the magnetic field's strength and morphology between the MIFEDS and no-MIFEDS experiments is also evident in the proton-radiography data, reconstructed path-integrated magnetic fields, and magnetic-energy spectra obtained at the later times (see Figure \ref{fig:PRAD4}, right panel). Intriguingly, even though the correlation length is similar, the characteristic value of the rms magnetic-field strength is somewhat reduced at late times compared to earlier ones in both MIFEDS and no-MIFEDS experiments: $B_{\rm rms} \approx 50$ kG at 38.7 ns (as compared with $B_{\rm rms} \approx 100$ kG at 31.2 ns). A plausible explanation for this observation is the decay of the turbulent kinetic energy by this stage of the interaction-region plasma's evolution (which has been seen in simulations of similar experiments~\cite{Bott2021}).

In summary, the proton radiography data confirm that the magnetic field in the interaction-region plasma post-amplification is not significantly altered by the MIFEDS in spite of much stronger seed magnetic fields and somewhat distinct initial flow dynamics in the interaction-region plasma.

\section{Discussion} \label{sec:discussion}

In the experiments described above, we have found that introducing a magnetic seed field $B_0$ into a turbulent, $\mathrm{Rm}$-supercritical laser-plasma that is six times larger than the self-generated Biermann seed field  does not lead to larger values of $B_{\rm rms}$ post-amplification; instead, $B_{\rm rms}$ seems to meet an inherent upper bound. This is inconsistent with the magnetic field being dynamically insignificant; in resistive magnetohydrodynamics (MHD), which is an appropriate model for the collisional laser-plasmas present in the experiment, the evolution of a dynamically insignificant field is linear, and thus is proportional to $B_0$. We conclude that $B_{\rm rms}$ must be dynamically significant with respect to turbulent motions in the interaction-region plasma. This result is perhaps surprising, given the ${\sim}3\%$ value of the magnetic to turbulent-kinetic energy ratio. However, as was discussed in the Introduction, it is consistent with the results of earlier laboratory experiments~\cite{Tzeferacos2018,Bott2021}. Further, periodic-box MHD simulations of fluctuation dynamo with similar Rm and Pm values find that the magnetic field begins to back-react on the turbulent motions once $\epsilon_{\rm B}/\epsilon_{\rm K,turb} \gtrsim 1\%$, although the saturation value of $\epsilon_{\rm B}/\epsilon_{\rm K,turb}$ at comparable $\mathrm{Rm}$ and $\mathrm{Pm}$ tends to be somewhat larger ($\epsilon_{\rm B}/\epsilon_{\rm K,turb} \approx 8\%$~\cite{Seta2020,Schekochihin2004}) than the reported experimental values. 

There are two possible explanations for the latter discrepancy: first, that field growth has fully saturated in the experiment at a smaller energy ratio (for reasons that may have to do, for example, with the experimental plasma being neither fully incompressible~\cite{Federrath_2011,Chirakkara_2021} nor spatially homogeneous and periodic, as the numerical one is); second, that an insufficient number of driving-scale eddy-turnover times have passed in the experiment for the dynamo to have saturated. The second possibility, which might naively seem counterintuitive as it would require identical transient magnetic-field strengths to be reached starting with two different seed fields over the same period of time, cannot in fact by ruled out or corroborated by our experimental results. This is because the initial field in the MIFEDS experiment ($B_0 \sim 60 \,  \mathrm{kG}$), while larger than in the no-MIFEDS one ($B_0 \sim 10 \, \mathrm{kG}$), is still small enough for its amplification to start in the kinematic phase of dynamo action; in both experiments, the magnetic field first grows exponentially fast at a rate $\gamma_{\rm kin}$ to a dynamical strength $B_{\rm nl}$ over a very short time ($t_{\rm kin}$), and then spends most of the time being amplifying further in the nonlinear, secular regime. It is then natural that measurements at a time interval $\Delta t \sim 6$ ns $\gg t_{\rm kin}$ after the jet collision would find the same state. Based on previous time-resolved measurements of the magnetic field~\cite{Bott2021}, we estimate that $\gamma_{\rm kin} \sim 1.8 \times 10^9 \, \mathrm{s}^{-1}$, $B_{\rm nl} \sim 86$ kG, and so $t_{\rm kin} \sim 1.2$ ns ($\Delta t-t_{\rm kin} \sim 4.8$ ns) in the no-MIFEDS experiments and $t_{\rm kin} \sim 0.2 \, \mathrm{ns}$, ($\Delta t-t_{\rm kin} \sim 5.8$ ns) in the MIFEDS ones. In both cases, $\Delta t - t_{\rm kin}$ is comparable to ${\sim}$1-2 driving-scale eddy turnover times $\tau_{\rm eddy} (\sim 4 \, \mathrm{ns})$, whereas in periodic-box simulations, saturation of the dynamo takes ${\sim}3$-$5\tau_{\rm eddy}$ after the beginning of the nonlinear dynamo regime — a somewhat longer period than our experiment lasts. We remain uncertain about which possibility is the correct one; more experiments with time-resolved measurements over a longer period and/or with larger seed fields, closer to the current level achieved at the end of the experiment, will be needed before either possibility can be corroborated definitively.

In summary, our results support a key prediction of theoretical dynamo theory: that increasing the initial seed field's strength does not lead to larger characteristic magnetic-field strengths post-amplification in a turbulent, magnetised fluid. More generally, it also suggests that in turbulent, $\mathrm{Rm}$-supercritical plasmas, magnetic fields will tend to undergo spontaneous amplification, and become dynamically significant. In addition to astrophysical applications, this conclusion is relevant to recent inertial-confinement fusion efforts that are attempting to leverage pre-imposed magnetic fields to control heat transport~\cite{Walsh2020}; if turbulence-generating fluid instabilities such as the Rayleigh-Taylor instability are also present in such systems, it is possible that dynamo-amplified magnetic fields could play a role in the subsequent dynamics.

\begin{acknowledgments}
The research leading to these results has received funding from the European Research Council under the European Community’s Seventh Framework Programme (FP7/2007-2013)/ERC grant agreements no. 256973 and 247039, the U.S. Department of Energy (DOE) National Nuclear Security Administration (NNSA) under Contract No. B591485 to Lawrence Livermore National Laboratory (LLNL), Field Work Proposal No. 57789 to Argonne National Laboratory (ANL), Subcontracts No. 536203 and 630138 with Los Alamos National Laboratory, Subcontract B632670 with LLNL, and grants No. DE-NA0002724, DE-NA0003605, and DE-NA0003934 to the Flash Center for Computational Science, DE-NA0003868 to the Massachusetts Institute of Technology, and Cooperative Agreement DE-NA0003856 to the Laboratory for Laser Energetics University of Rochester. We acknowledge support from the U.S. DOE Office of Science Fusion Energy Sciences under grant No. DE-SC0016566 and the National Science Foundation under grants No. PHY-1619573, PHY-2033925 and PHY-2045718. We acknowledge funding from grants 2016R1A5A1013277 and 2020R1A2C2102800 of the National Research Foundation of Korea. Support from AWE plc., the Engineering and Physical Sciences Research Council (grant numbers EP/M022331/1, EP/N014472/1, and EP/R034737/1) and the U.K. Science and Technology Facilities Council is also acknowledged. The authors thank General Atomics for target manufacturing and R\&D support, funded by the U.S. DOE NNSA in support of the NLUF program through subcontracts 89233118CNA000010 and 89233119CNA000063.
\end{acknowledgments}

\section*{Data Availability Statement}

The data that support the findings of this study are available from the corresponding author upon reasonable request.

\nocite{*}
\bibliography{aipsamp}

\end{document}